# Cutting "lab-to fab" short: High Throughput Optimization and Process Assessment in Roll-to-Roll Slot Die Coating of Printed Photovoltaics

Michael Wagner[1,2], Andreas Distler[2], Vincent M. Le Corre[2], Simon Zapf[2], Burak Baydar[2], Hans-Dieter Schmidt[2], Madeleine Heyder[2], Karen Forberich[1,2], Larry Lüer[1,2], Christoph J. Brabec[1,2], H.-J. Egelhaaf[1,2]

[1] Forschungszentrum Jülich GmbH, Helmholtz Institute Erlangen-Nürnberg for Renewable Energy (HI ERN), Dept. of High Throughput Methods in Photovoltaics, Erlangen, Germany

[2] Friedrich-Alexander-Universität Erlangen-Nürnberg, Materials for Electronics and Energy Technology (i-MEET), Erlangen, Germany

## Abstract

Commercialization of printed photovoltaics requires knowledge of the optimal composition and microstructure of the single layers, and the ability to control these properties over large areas under industrial conditions. While microstructure optimization can be readily achieved by lab scale methods, the transfer from laboratory scale to a pilot production line ("lab to fab") is a slow and cumbersome process: first, the difficulty of operating structure-sensitive methods in-line impedes proper microstructure characterization, and second, the processing-functionality relationship must be redetermined for every material combination as the results obtained by typical lab-scale spin-coating cannot be directly transferred to other coating methods. Here, we show how we can optimize the performance of organic solar cells and at the same time assess process performance in a 2D combinatorial approach directly on an industrially relevant slot die coating line. This is enabled by a multi-nozzle slot die coating head allowing parameter variations along and across the web. This modification allows us to generate and analyze 3750 devices in a single coating run, varying the active layer donor:acceptor ratio and the thickness of the electron transport layer (ETL). We use Gaussian Process Regression (GPR) to exploit the whole dataset for precise determination of the optimal parameter combination. Performance-relevant features of the active layer morphology are inferred from UV-Vis absorption spectra. By mapping morphology in this way, small undesired gradients of process conditions (extrusion rates, annealing temperatures) are detected and their effect on device performance is quantified. The correlation between process parameters, morphology and performance obtained by GPR provides hints to the underlying physics, which are finally quantified by automated high-throughput drift-diffusion simulations. This leads to the conclusion that voltage losses which are observed for very thin ETL coatings are due to incomplete coverage of the electrode by the ETL, which cause enhanced surface recombination.

# Introduction

Printed photovoltaics (PV) is on the brink of commercialization, with several companies already producing printed PV modules for different applications, ranging from large-area building-integrated PV to small-scale sensors.[i, ii] Record efficiencies have increased steeply over the past years, especially for organic solar cells (OSCs)[iii, iv] and modules.[v] The ongoing synthesis of new materials, in combination with the improved understanding of the relationship between microstructure formation and voltage losses, shows great potential to further increase performance.[vi, vii] High throughput methods, combined with machine learning and statistical analysis, have been established to accelerate the screening of these materials.[viii, ix, x, xi, xii] However, up to now, most of these studies have only been performed on small devices areas. For layer deposition, coating techniques such as spin coating[xiii] and doctor blading,[xiv, xv, xvi] as well as printing techniques, such as ink jet printing,[xvii] have been employed. To leverage the exploration rate of multi-dimensional parametric spaces further, 2D combinatorial libraries in a single substrate have been produced by applying perpendicular gradients of annealing temperature and film thickness.[xvi, xviii] These techniques allow high throughput at low material consumption and are easy to automate but are limited to small device areas. Furthermore, the conclusions derived cannot be directly transferred to large-scale production methods: for instance, different drying conditions can lead to different morphologies, and there are stricter requirements on the environmental compatibility of the solvents that are used. There are only few reports in which industrially compatible methods have been employed, for instance the determination of the optimum active layer properties by generating 1D gradients of active layer thickness or donor:acceptor (D:A) ratio in roll-to-roll (R2R) coating of organic photovoltaic (OPV) devices.[xix, xx] Combining these production-compatible screening methods with the extremely high output of the 2D lab scale methods will enable the exploration of the large data space of correlated parameters in the production of printed PV modules and will allow to reduce the time for transferring recipes from lab to fab to a minimum.

In this work, we therefore present a novel high-throughput method for screening materials and optimizing coating processes which combines coating on R2R production equipment with the generation of 2D combinatorial patterns and thus provides rapid parameter variation under industrially relevant conditions. For this purpose, we use our R2R coating line (Figure S2) for slot-die coating and laser patterning. Using the D:A system poly-(3-hexylthiophene-2,5-diyl): 5, 5'- [(4, 9- dihydro- 4, 4, 9, 9- tetraoctyl- s- indaceno[1, 2- b:5, 6- b'] dithiophene- 2, 7- diyl) bis(2, 1, 3- benzothiadiazole- 7, 4- diylmethylidyne) ] bis[3- ethyl- 2- thioxo-4-Thiazolidinone (P3HT:o-IDTBR) as a benchmark,[xxi] we validate the potential of this method to simultaneously vary the electron transport layer (ETL) thickness, in cross-web direction, and the D:A ratio of the bulk heterojunction, along web direction, thus obtaining 3750 individual devices in a single coating run. To put this number into context, gathering this much information by conventional, non-automated experiments with an average number of 20 substrates per day would take around half a year, with an inevitable variation of experimental conditions. In addition, conventional experiments require several devices for each parameter set to obtain a statistically relevant result, whereas in our combinatorial method this is no longer required due to the clear trends that can be observed. We will thus be able to skip most of the lab scale optimization procedures by directly moving from the spin-coater to the R2R production line.

In order to extract as much information as possible from the wealth of data provided by high-throughput experimentation, several publications report on the application of statistical methods and machine learning for the analysis of device performance and additional spectroscopic information.[xx, xviii]

In our work, we employ Gaussian process regression for statistical analysis of the huge amount of data obtained from the 2D experiment, which results in a dramatic reduction of uncertainty for determining the optimal process parameters. In combination with spectroscopically derived performance-relevant morphology features,[xxii] we can assess process homogeneity. The level of detail extracted from optical spectra allows us to identify if individual process parameters are gradually changing during the experiment. Moreover, connections between morphology, processing parameters, and performance hint at the underlying device physics. Finally, a novel method for automated, high throughput drift-diffusion simulations is presented which provides deeper insight into the parameters affecting the efficiencies of the resulting devices.

## Results and discussion

In order to demonstrate the potential of our 2D combinatorial approach, we have chosen organic solar cells (OSC) with the architecture PET/IMI/SnO$_2$/P3HT:o-IDTBR/PEDOT:PSS/AgNW, where PET stands for poly(ethyleneterephthalate), IMI for the transparent indium tinoxide/silver/indium tinoxide electrode, SnO$_2$ for tin oxide, PEDOT:PSS for poly-(2,3-dihydrothieno-1,4-dioxin)-poly-(styrenesulfonate), and AgNW for the silver nanowire electrode. We have further chosen the thickness of the electron transport layer (ETL) and the donor:acceptor (D:A) ratio in the absorber layers as the two parameters to vary because of their critical influence on the properties of the resulting OSC.

To optimize both parameters in a single coating run, a 15 m PET/IMI roll was roll-to-roll laser structured with a pattern and markers designed for fast and reliable measurement of the resulting solar cells (see Figure S3). The thickness of the ETL was varied perpendicular to the coating direction and the D:A ratio was varied along the coating direction. Figure 1a shows a schematic drawing of the experimental setup. For ETL thickness variation, five stripes of SnO$_2$ suspension of different concentrations were coated in parallel on coating station 1 (SD1) with a specially designed slot-die, with which up to ten separate reservoirs can be fed by separate channels (Figure S2). Based on our previous experience with the suitable thickness values of SnO$_2$, the concentrations of the suspensions were chosen so that the nominal dry film thickness varies on a logarithmic scale from 1.4 nm to 110 nm (1.4nm, 4.1nm, 12.2nm, 36.6nm, 110nm). Photographs of the coating can be found in Figure 1b and Figure 1 c.

P3HT and o-IDTBR inks were prepared individually and filled into four syringes (two syringes for each material) at coating station 2 (SD2). Here, we used a slot die with three inlets and one continuous reservoir (see Figure S2). The two inlets on the outer sides of the die were used for the semiconductor ink, whereas the middle inlet was used for degassing to eliminate air bubbles. This setup provides an extensive mixing of the two components inside the reservoir, as opposed to mixing only in the meniscus. The D:A variation was started with coating only the ink containing the o-IDTBR. After the steady-state of o-IDTBR coating had been obtained, the supply of o-IDTBR ink was stopped by turning off the corresponding syringe pump and the supply of P3HT ink was started by switching on the other syringe pump. At that point in time, the reservoir and the channels in the die are still filled with o-IDTBR ink. This ink is mixed with the P3HT ink, resulting in a gradual increase of the P3HT content and thus a change of the D:A ratio in the printed film, ranging from 0:1 at the start of the experiment to 1:0 at the end of the experiment. The fact that the observed change in D:A ratio is gradual indicates turbulent mixing, since a plug flow would result in an abrupt change from 100 % o-IDTBR to 100 % P3HT.

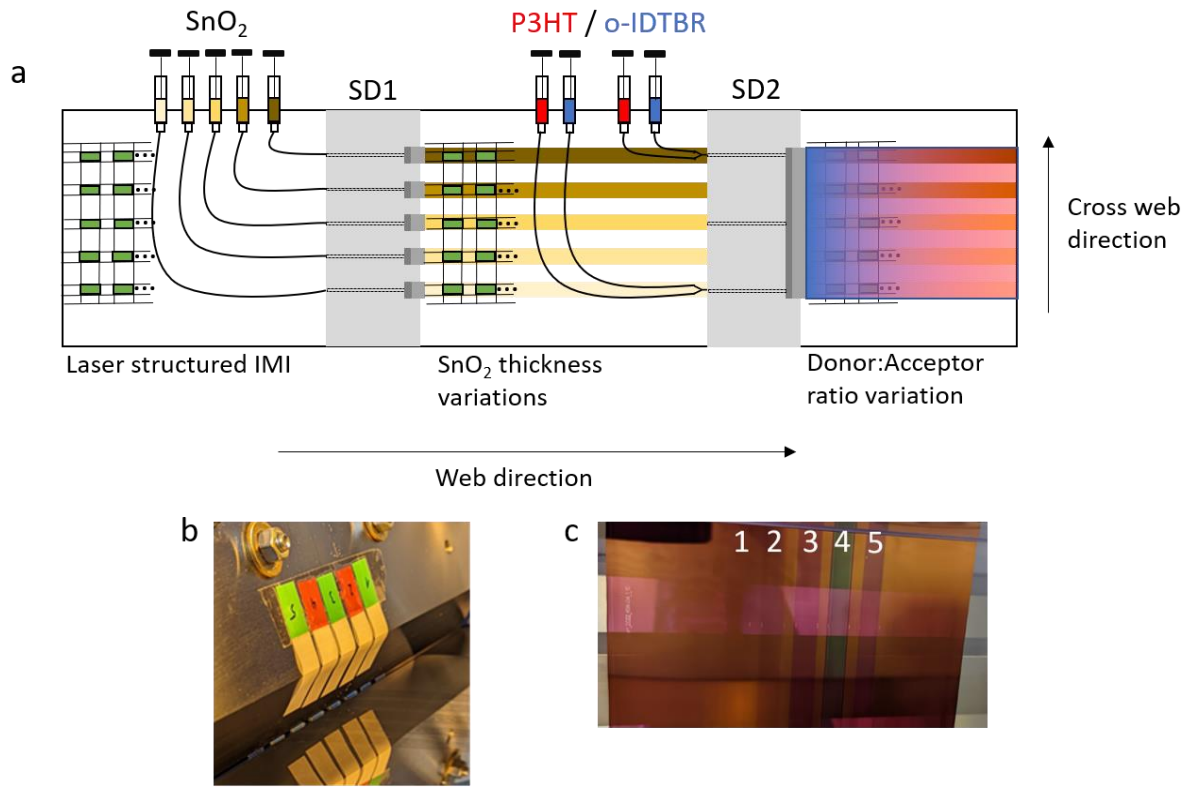

Figure 1: (a) Schematic drawing of the high throughput experiment. SnO$_2$ dispersions with five different concentrations were coated with a multi nozzle slot die onto a laser-patterned IMI substrate in five parallel stripes at Slot Die Coating station 1 (SD1) to provide ETLs of different thicknesses. Subsequently, the bulk heterojunction layer was coated at Slot Die station 2 (SD2), varying the donor:acceptor ratio along the printing direction from D:A = 0:1 to 1:0 . The coating of PEDOT:PSS and AgNW as well as the final laser patterning step are not included in this schematic. (b) Slot-die coating of ETL thickness variations, where the numbers 1-5 label the different variations of the ETL thicknesses (1=1.4 nm, 2=4.1 nm, 3=12.2 nm, 4=36.6 nm, 5 =110 nm calculated dry film thickness). (c) Printed active layer on top of SnO$_2$ stripes (vertical), photograph taken inline (horizontal stripe belongs to a structure behind the substrate).

The D:A ratio was determined at every position of the web by recording absorbance spectra at intervals of 4 mm during the coating run, on each of the 5 stripes of the SnO$_2$ variation, thus providing the spatial development of the mass fractions of donor and acceptor (Figure 2). Figure 2 also shows the photographs of the active layers at three exemplary positions, corresponding to the P3HT:o-IDTBR ratios of 0:1, 1:1 and 1:0, along with the corresponding absorbance spectra.

Figure 2With these data, we can assign a D:A ratio to the IV measurements of each individual cell. Further data about the spectral data processing and analysis can be found in the SI (Eq. S1, Eq. S2 and Figure S1). In the remainder of the manuscript, the D:A ratio will be given in terms of the acceptor weight fraction w$_A$ since the latter varies between values of 0 and 1 and is therefore more convenient for statistical analysis.Figure 3

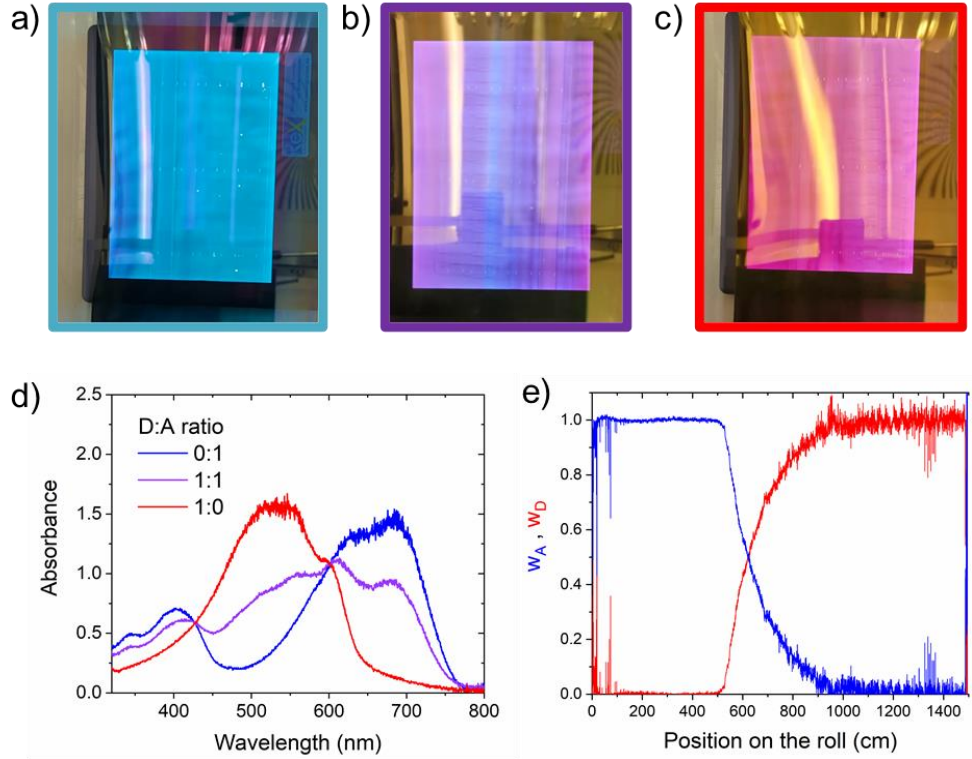

*Figure 2: Inline photographs (top row) of the active layer with D:A ratios of 0:1 (a), 1:1 (b), and 1:0 ratio (c). d) Corresponding absorbance spectra for the different D:A ratios. e) Mass fraction of donor ($w_D$) and acceptor ($w_A$) during the printing process as calculated from the absorbance spectra, plotted against the position on the substrate.*

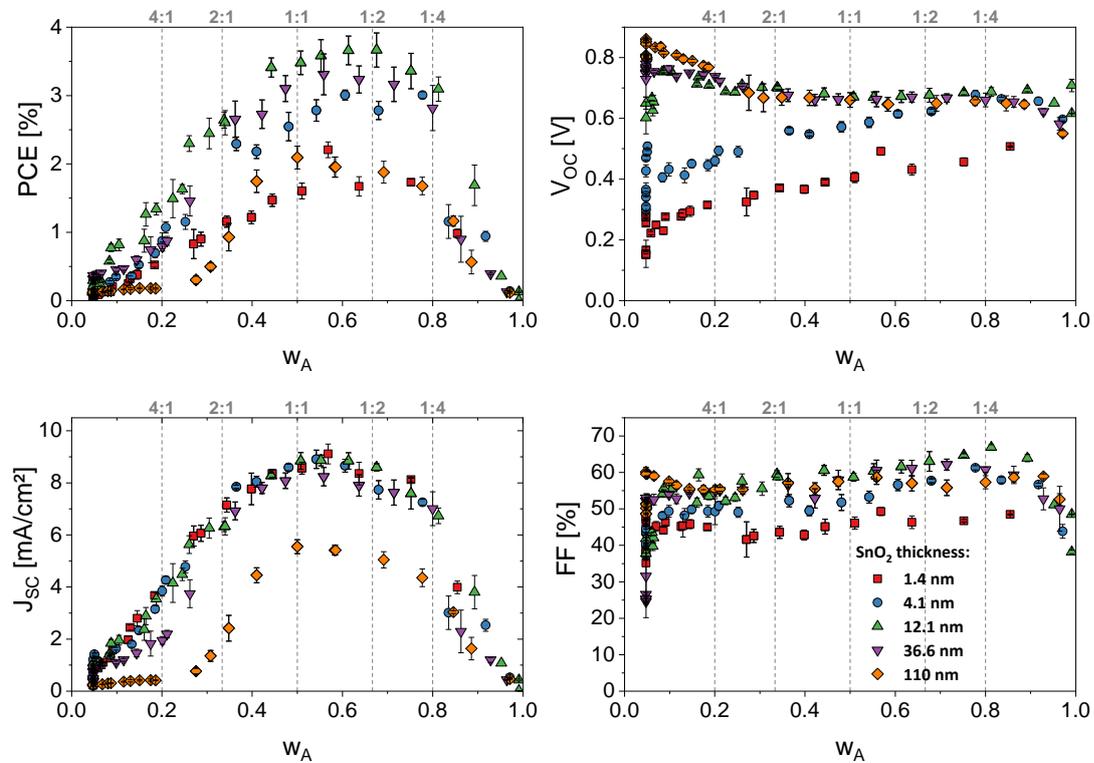

*Figure 3: Electrical key parameters (PCE – photoelectrical conversion efficiency, $V_{oc}$ – open circuit voltage, $J_{sc}$ – short circuit current, FF – fill factor) of the high throughput experiment, with different ETL thicknesses (1.4, 4.1, 12.1, 36.6, 110 nm) displayed in different colors and symbols. The x-axes show the mass fraction $w_A$ of the acceptor. The dashed lines are guides to the eye to the corresponding donor:acceptor ratios.*

After completing the solar cells by coating HTL and AgNW electrodes, the web is cut into sheets of 18 cm in length, which are divided into 50 solar cells by laser patterning of the top electrode. The resulting 3750 cells are subsequently characterized under a sun simulator with respect to their current density-voltage (JV) characteristics, using fully automated equipment consisting of measurement board, source measurement unit and multiplexer (for details see Experimental Part).

Figure 3 shows the electrical key parameters of the two-dimensional parameter variation, with **$w_A$** ranging from 0 to 1 and the SnO$_2$ layer thickness ranging from **$d_{ETL}$** = 1.4 nm to 110 nm. For this evaluation, it was assumed that the variation in $w_A$ within one sheet (18 cm in coating direction) is negligible and so every data point is the average of 10 measured cells, which are located on the same sheet.

It is immediately obvious that intermediate acceptor weight fractions **$w_A$** and SnO$_2$ thicknesses **$d_{ETL}$** result in the best performance, mostly due to a maximum in short circuit current ($J_{sc}$). The datapoint with the highest efficiency is obtained for **$w_A$** = 0.67, or a D:A ratio of 1:2, and a SnO$_2$ layer thickness of 12.1 nm. However, the efficiency measurement of a single device is subject to statistical uncertainty (see error bars in Figure 3). In fact, considering the 95% confidence interval given by the double of the error bars, the optimal D:A ratio could be anywhere between 0.45 and 0.82, if the single data points were used to identify the optimum. Moreover, the sampling density along the cross-web direction is scarce (only 5 different SnO$_2$ thicknesses), so that the optimal thickness might fall in between the experimentally realized values.

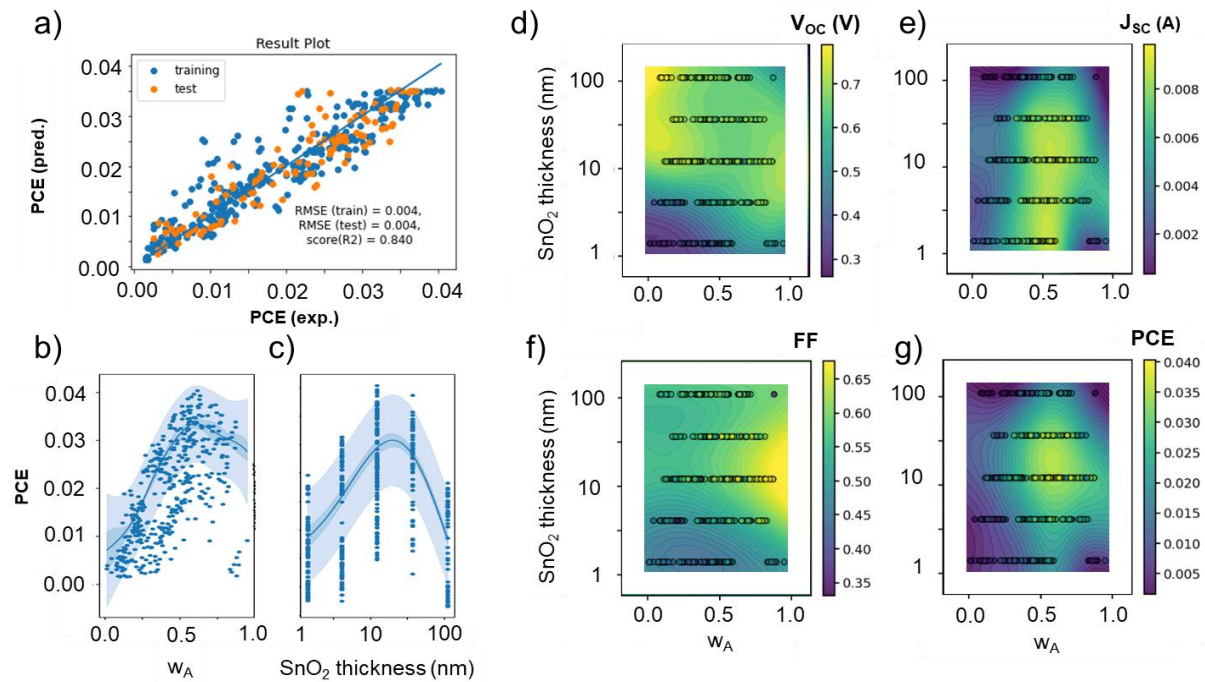

*Figure 4: Gaussian Process Regression (GPR) to access device photophysics. a) result plot showing predicted against experimental PCE. Training and test datasets given as blue and orange symbols, respectively, with the corresponding root mean square errors (RMSE) indicated as inset, where also the R2 score for a 5-fold cross-validation is given. b) and c): one-dimensional intersections through the approximate objective function $PCE^{GPR} = f(w_A, log(d_{ETL}))$ (blue solid line), light blue and dark blue areas: 95% confidence intervals for the uncertainty of a single prediction and uncertainty of the mean, respectively. d-f): approximate objective function $T^{GPR} = f(w_A, log(d_{ETL}))$ (coloured hypersurface), where T ={**$V_{OC}$, $J_{SC}$, FF, PCE**}, respectively.*

The large number of available data points suggests a regression analysis to reduce the uncertainty and to predict the optimal parameter combination by exploiting the information from the whole dataset, rather than just considering single data points.

To this end, we use Gaussian Process Regression (GPR), a probabilistic method for fitting a general, non-linear objective functions PCE ($V_{oc}$, $J_{sc}$, FF) = f($w_A$, $d_{ETL}$) to the dataset. The quality of the fit is assessed using a so-called result plot (Figure 4a), in which the predicted PCE values are displayed against the experimental ones. To avoid overfitting, we exclude 25% of the data from the training dataset (blue symbols) and use them to test the objective function on unseen experimental points ("test dataset", orange symbols); Figure 4a shows that test and training data sets both show the same root mean square error (RMSE) of $4*10^{-3}$, evidencing that there is no overfitting. The uncertainty of the objective function is much smaller than the uncertainty of a single observation, as obvious from comparing dark and light blue areas, respectively, in Figure 4b and c. Especially, Figure 4c shows that the optimal $SnO_2$ thickness should be between the experimentally realized values of 12.1 and 36.6 nm. Using this information, we infer the optimal parameter combination as $w_A$ = 0.59 ± 0.04, corresponding to a D:A ratio of 1:1.44, and $d_{ETL}$=17 ± 4 nm (see ESI, Part 5). Thus, the uncertainty along the $w_A$ direction is much smaller than the one achievable considering individual measurements, while along the $d_{ETL}$ direction, the uncertainty is even smaller than the spacing of the experimental data points, clearly highlighting the strength of the high-throughput method.

Another important aspect of probabilistically derived objective functions is the fact that they give insight into the underlying device photophysics. Figure 4d, e, f, and g show objective functions predicting $V_{OC}$, $J_{SC}$, $FF$, and $PCE$, respectively, as function of $w_A$ and $d_{ETL}$, displayed as two-dimensional hypersurfaces with the objective function value colour coded. Strikingly, all of the objective functions are non-orthogonal, meaning that extrema along one dimension depend on the value of the second dimension. It follows that they cannot be factorized into one-dimensional functions. Naively, one would have assumed that a bulk property such as $w_A$ and an interface property such as $d_{ETL}$ are independent in their influence on the device parameters. This clearly highlights the importance of a complete scan of the parameter space, while "Edisonian experiments" (varying one parameter at a time) implicitly assume orthogonality and would have missed this important, physics-related property.

Figure 4d shows that the trend for $V_{OC}$ as function of $w_A$ depends on $d_{ETL}$: if the ETL is thin then acceptor-rich blends are needed to get high $V_{OC}$. In contrast, if the ETL is thick then the highest $V_{OC}$ values are observed for donor-rich blends, which is generally the expected behaviour for a variation of the D:A ratio.[xxii] Similar trends are observed for FF, see Figure 4f. Interestingly, $J_{sc}$ does not show such a trend, maximum $J_{sc}$ always occurring for similar $w_A$ values, independent of $d_{ETL}$.

In the following, we will try to identify the underlying physical reasons for the observed interdependence of ETL thickness and D:A ratio with respect to their effect on $V_{oc}$. A possible scenario for the need for acceptor-rich blends in the case of thin ETL could be incomplete coverage of the IMI substrate by the ETL, causing direct contact between active layer and IMI. This will cause hole transfer into IMI and subsequent recombination with the electrons from the acceptor, under two conditions: the donor phase is in direct contact with the IMI and the hole density in the donor phase is sufficiently high to cause recombination with majority electrons. The first condition would explain why donor-rich blends have reduced $V_{oc}$ (reduction of QFLS due to strong surface recombination), while the second condition would explain why $J_{sc}$ is not affected (no hole accumulation under extraction conditions). Another possible scenario would be a different active layer morphology for different $d_{ETL}$ values, either caused directly by the $SnO_2$ or indirectly correlated, e.g., by undesired gradients in cross-web direction y of active layer processing parameters.

To assess the role of the latter scenario, we have applied a method presented by us recently,[xxii] which is based on Spano's model of weak H-aggregates,[xxiii, xxiv] to analyse the active layer morphology and its effect on $V_{oc}$ by spectral decomposition of the UV-Vis spectra (for details, see Figure S4 and ref. xxii). In order to make sure that predictors are only included into the final regression if they contribute significant additional explanation of variance, we embedded a maximum relevance/minimum redundancy feature selection scheme into GPR (mRMR-GPR). For a detailed explanation, see ESI, Figure S5. Figure 5a shows that on the basis of the bulk heterojunction morphologies extracted from the UV-Vis spectra, we can predict $V_{OC}$ with an RMSE of 67 mV. In ESI, Figure S5, we show that the total absorption $a_{tot}$, the donor exciton energy $c_D$ and its bandwidth $b_D$ are the most relevant predictors for $V_{OC}$. However, if we include the ETL thickness in the feature list, the prediction of $V_{OC}$ is significantly improved, see Figure 5Figure 5b. In this case, the RMSE is only 36 and 37 mV for the training and test datasets, respectively. This means that knowledge of the ETL thickness improves the prediction of $V_{OC}$. This result shows that the observed dependence of $V_{OC}$ on $d_{ETL}$ cannot be explained by active layer morphology alone, pointing to the interface as a decisive influence on $V_{oc}$.

Before we turn to the elucidation of the role of the interface on $V_{oc}$, we would like to mention an additional benefit of the feature selection scheme employed here, which turns out to be a powerful tool for identifying processing instabilities.

The feature selection scheme finds redundancy between $a_{tot}$, $b_D$ and $d_{ETL}$ which means that there is a correlation between these morphological features and the $SnO_2$ thickness. This is corroborated in Fig. S8 (see ESI), where we show that $a_{tot}$ and $c_A$ have indeed the highest correlation of all morphology features with the y (cross-web) position. Since $d_{ETL}$ is also varied along y, and $d_{ETL}$ influences $V_{OC}$, a correlation of morphology with $V_{OC}$ must result. The variation of $a_{tot}$ and $c_A$ along y is probably due to an undesired small gradient of AL process conditions along the y (cross-web) direction. A variation of $a_{tot}$ along y may be due to incomplete mixing of the two components inside the reservoir and before exiting the slot die. Furthermore, a variation of the donor exciton energy $c_D$ and bandwidth $b_D$ is known to depend on annealing, so that a small temperature gradient in the annealing oven may be one of the reasons causing this morphology gradient. As shown in Fig. S9c, the acceptor exciton energies are clearly increased at the edges of the web along the whole web, which speaks against a direct correlation with the ETL thickness, as this would entail a monotonous, rather than a symmetric trend along the cross-web direction.

Elucidating the effect of the ETL thickness on $V_{oc}$ requires insight into its electronic properties as a function of its thickness. However, experimentally assessing coverage of substrates by ultrathin buried layers is very difficult, for which reason indirect methods are being deployed. These methods detect incomplete coverage by the consequences on the energy levels exerted by direct contact of the active layer with the electrode, which is expected to influence the effective work function of the ETL. In order to extract the interfacial band structure from current-voltage (IV) curves, we have used automated drift-diffusion fitting with a Bayesian optimization method that can, in principle, be used with any numerical model. In our case, we have used our homemade Bayesian optimization package BOAR (Bayesian Optimization for Automated Research) and the open-source software SIMsalabim[xxv] to calculate IV curves by solving the one-dimensional drift-diffusion model assuming virtual semiconductors. The details of the method are described in part 6 and 7 of the ESI.

The ability to fit a large number of parameters makes BOAR apt to distinguish between the two scenarios mentioned above, because we can vary the parameter sets describing the active layer and the ETL simultaneously. In particular, we have used the bulk trap density, the bimolecular recombination coefficient, the electron and hole mobility and the charge generation rate $G_{ehp}$ as parameters describing the properties of the active layer, and the work function $W_L$ of the ETL as well as the surface trap density between ETL and active layer as parameters describing the ETL. The work function of the ETL was varied under the assumption that its effective value will lie in between the corresponding values of ITO and SnO$_2$ in the case that the coverage of SnO$_2$ on the IMI substrate is incomplete. Series and parallel resistance were also included in the fit parameters. The thickness of the ETL ($d_{ETL}$) is set to the value which corresponds to the position of the respective cell on the sheet.

The full set of resulting parameters is shown in the ESI, Table S1 and Figure S10. The only one of the parameters referring to the active layer morphology that shows a clear trend with the D:A ratio is the charge generation rate $G_{ehp}$ (Figure 5e). This is expected because the amount of o-IDTBR controls the amount of absorptance in the red spectral region and thus should enhance $G_{ehp}$. Furthermore, we did not find any parameter referring to active layer morphology that exhibits a clear dependence on $d_{ETL}$. In contrast, looking at the parameters describing the ETL, there is a clear dependence of the ETL work function $W_L$ on $d_{ETL}$, see Figure 5f. The thinnest ETL has a significantly higher work function than the thicker ones, and the difference in work function increases from 0.2 to 0.4 V when going to donor-rich blends. This finding fully confirms the scenario of incomplete coverage being the reason for the $V_{OC}$ losses in donor-rich blends and thin ETL.

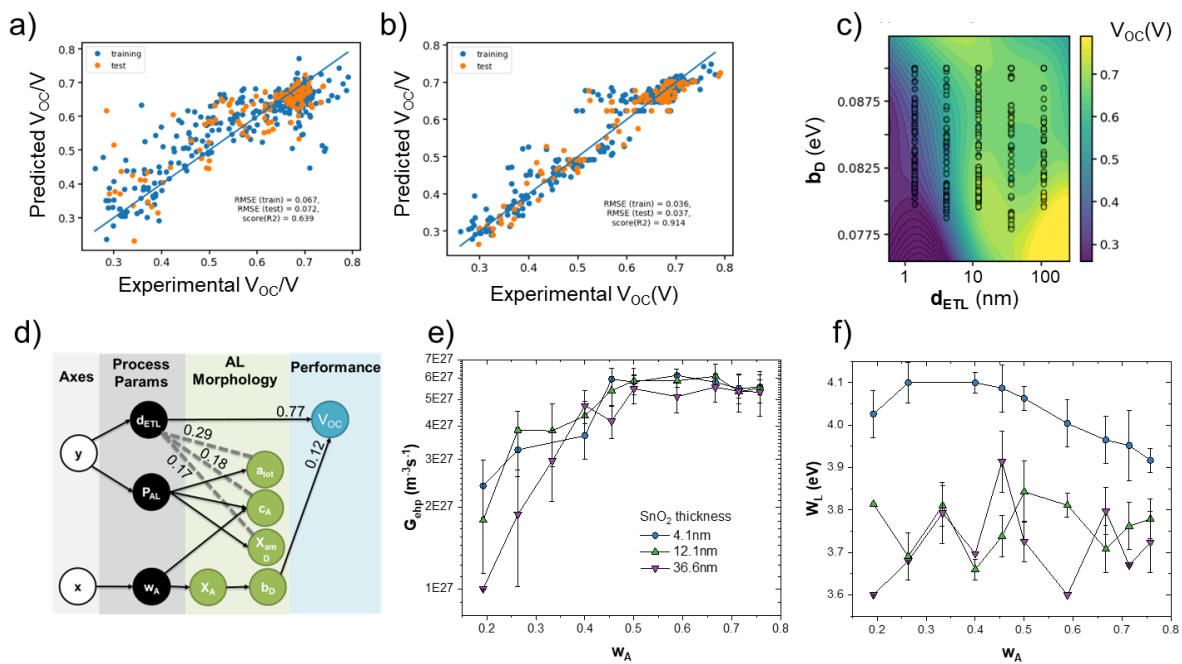

Figure 5: a) Result plot for a GPR to predict V$_{OC}$ only from morphological predictors extracted from UV-Vis absorption spectra; b) Same as a) but including the ETL thickness in the list of predictors. C) Objective function $V_{OC} = f(d_{ETL}, b_D)$ comprising only non-redundant predictors (hypersurface with $V_{oc}$ value given as color bar). Symbols: experimental data points. d) Knowledge graph constructed from the mRMR-GPR runs, showing direct causations (full black lines) and non-causal pathways (correlations, gray dashed lines). e) Effective charge generation rate $G_{ehp}$ obtained from fitting the drift-diffusion simulations to the measured JV curves. f) Effective ETL work function $W_L$ obtained by the same method as in e).

# Conclusion

We have used a 2D combinatorial approach with five different ETL thicknesses and donor:acceptor ratios ranging from 100% acceptor to 100% donor for the fabrication of organic solar cells in a single slot die coating experiment on industrial roll-to-roll equipment. The experiment resulted in a large data set of JV curves and UV-Vis absorption spectra for 3750 devices, corresponding to 3750 different combinations of ETL thickness and D:A ratio. The exceptional quality of this data is evidenced by the Gaussian Process Regression (GPR) prediction of the optimum efficiency that is found for an ETL thickness of 17 ± 4 nm and a D:A ratio of 1:1.44 with an uncertainty one order of magnitude lower than the uncertainty of a single measurement. Statistical analysis of the large number of high-quality data reveals non-orthogonal dependencies of $V_{oc}$ on ETL thickness and D:A ratio, observing higher $V_{oc}$ losses for high D:A ratios when ETLs are thin. We employ a physics-informed approach to elucidate the reasons for this behavior. By using spectral fitting of UV-Vis transmittance spectra combined with a maximum relevance/minimum redundancy feature selection scheme embedded into GPR and drift-diffusion fitting of IV curves, we identify incomplete coverage of the IMI electrode by the thinner of the ETLs and consequently enhanced interface recombination of photogenerated charge carriers as the main reason for $V_{oc}$ loss.

In addition, redundancies between exciton properties and ETL thickness reveal unintended and otherwise unrecognized spatial gradients in processing conditions, probably caused by insufficient mixing of donor and acceptor inks as well as gradients of annealing temperature across the web.

In essence, combinatorial device preparation on R2R production equipment does not only allow to reduce the effort for lab-scale optimization of solar cells to a minimum. Owing to the large amount of high-quality data produced by this method, statistical analysis provides hidden parameters, which reveal not only interdependencies of processing parameters in their effects on device key performance indicators but also otherwise undetected fluctuations in processing conditions. Complementing statistical analysis with physics informed methods allows us to obtain an understanding of production failures on the device level.

# Experimental details

## Materials

The substrates utilized in this work are based on heat-stabilized polyethylene terephthalate (PET) (DuPont Teijin Films, Melinex® ST504) with transparent conductive coatings of ITO–Ag–ITO (IMI) and were purchased from OPVIUS. The active materials poly(3-hexylthiophene) (P3HT) and (5Z,5'Z)-5,5'-((7,7'-(4,4,9,9-tetraoctyl-4, 9-dihydro-s-indaceno[1,2-b:5,6-b']dithiophene-2,7- diyl)bis(benzo[c][1,2,5]thiadiazole-7,4-diyl))bis(methanylylidene bis(3-ethyl-2-thioxothiazolidin-4-one))) (O-IDTBR) were purchased from OPVIUS and Nano-C, respectively. The solvents used to dissolve the active layer materials were o-xylene (o-XY, Sigma-Aldrich) and 1-methylnaphthalene (1-MN, Merck). The charge transport layers, tin oxide ($SnO_2$, N31) and poly(3,4-ethylenedioxythiophene)-poly(styrenesulfonate) (PEDOT:PSS) (HTL Solar) were purchased from Avantama AG and Heraeus, respectively. Finally, water-based silver nanowire (AgNW) ink, with NWs of 25 nm diameter, was purchased from Zhejiang Kechuang Advanced Materials Technology Co., Ltd.

## Roll-to-Roll machines

The roll–to-roll (R2R) pilot coating machine (Grafisk Maskinfabrik, Denmark) comprises three slot-die coating stations. Two of them are equipped with 2 m long hot air ovens for drying and annealing of the deposited films, the third station (used for semiconductor coating) has a heating mat installed for this purpose, directly after the coating station. Stainless steel slot-dies were used for coating of all layers. The slot dies are equipped with shims to guide the ink and define the coating width. Syringe pumps and PTFE pipes are employed for pumping the ink into the slot dies. For the two syringes that are supplying the ink for the absorber layer, tubes and slot-die were heated to avoid precipitation of the semiconductors in the ink feeding systems. In order to record UV/Vis absorption spectra inline during the coating process over the whole width of the coating web, a fiber spectrometer (Ocean Optics Fame-S-UV-VIS-ES) with a movable fiber was installed after the heating mat.

For separating the coated layer stacks into cells, laser ablation was employed, using an LS-6KP4P520 R2R laser patterning machine (LS Laser Systems GmbH). This machine comprises the ultrafast laser source Spirit 1040-8-SHG (Spectra Physics) emitting an SHG-generated center wavelength of 520 ± 3 nm with a pulse duration of greater than or equal to 350 femtoseconds. Maximum power of up to 4 W can be achieved at a pulse repetition rate of 500 kHz. The beam was scanned over the sample using a galvanometer scanner with an f-theta lens of focal length of 506 mm, achieving deflecting speeds up to 4 m/s. The R2R laser machine is equipped with an unwinder, vacuum table and rewinder. Two cameras are used for exact positioning of the laser beam, providing a precision of better than 100 μm.

Device fabrication

The devices of the architecture PET/IMI/SnO2/P3HT:O-IDTBR/PEDOT:PSS/AgNW were prepared by the combination of laser patterning and slot die coating. In a first step, the PET/IMI substrates are roll-to-roll patterned with a fs-laser (LS Laser Systems) with 350 fs pulse duration, 520 nm wavelength, and 0.40 J cm$^{-2}$ fluence to electrically separate the IMI bottom electrode into individual cells. The pattern is organized into sheets, where one sheet is 18 cm long and consists of 50 cells (see a sketch of the pattern in Figure S3). The laser-patterned PET/IMI substrates were cleaned with an air blade, a Teknek cleaning roller and finally, with microfiber tissue and toluene to get rid of the debris caused by laser patterning. All layers are deposited in ambient air by slot-die coating. First, the SnO$_2$ inks were slot-die coated in five parallel stripes of 1 cm width on the patterned substrate and subsequently annealed at 130 °C for 4min inside the hot air oven adjacent to the coating station. The active material solutions, P3HT and o-IDTBR were prepared separately (20 mg/ml for each) in o-Xylene:1-MN (19:1) and stirred over night at 80 °C. The two solutions were injected into a single slot die by separate syringe pumps. During coating, syringes, tubes, slot die, and backing roll were heated to 80 °C. Immediately after coating, the wet film was heated with a heating mat at 90 °C for 45 s. The hole transport layer, PEDOT:PSS, is coated at room temperature followed by an inline annealing step at 140 °C for 4 min. Subsequently, the AgNW top electrode is deposited and annealed for 2 min at 130 °C, which results in a sheet resistance of approx. 8 Ω sq$^{-1}$. SnO$_2$, active layer and HTL were coated at a web speed of 0.5 m/min, AgNWs were coated at 1 m/min. After coating the top electrode, a final laser patterning step was conducted with 0.18 J cm$^{-2}$ laser fluence to establish the matrix of 10 x 5 individual solar cells, each with an area of 0.1375 cm$^2$, on every substrate. During the laser patterning process, holes for registration were scribed into each sheet, so that the sheets can be accurately aligned in the measuring board during the subsequent characterization steps.

Characterization

The inline UV/Vis absorption measurements were performed using a fiber spectrometer from Ocean Insight (Fame-S-UV-VIS-ES) with an operating wavelength between 200 and 850 nm and a halogen and deuterium tungsten light source.

The current density-voltage (JV) characteristics of the solar cells are measured by using a source measure unit (Keysight B2901A, Keysight Technologies) and a class AAA solar simulator (LOT Quantum Design) providing AM 1.5 G illumination of 1000 W cm$^{-2}$. In order to be able to measure one hundred solar cells without changing the substrate, a custom-designed measuring board was used, which can be flooded with nitrogen (10 coated stripes (y-direction) and 10 cells (x-direction)). Switching between cells is achieved by a custom-designed multiplexer unit.

# Acknowledgments

The authors acknowledge the 'Solar Factory of the Future' as part of the Energy Campus Nuremberg (EnCN), which is supported by the Bavarian State Government (FKZ 20.2-3410.5-4-5). The authors also acknowledge funding from the European Union's Horizon 2020 research and innovation program under grant agreement No. 952911 ("BOOSTER") and 101007084 ("CITYSOLAR"). H.-J. E. and C.J.B. acknowledge funding from the European Union's Horizon 2020 INFRAIA program under Grant Agreement No. 101008701 ('EMERGE'). Part of this work has been supported by the Helmholtz Association in the framework of the innovation platform "Solar TAP".

Michael Wagner[1,2], Andreas Distler[2], Vincent M. Le Corre[2], Simon Zapf[2], Burak Baydar[2], Hans-Dieter Schmidt[2], Madeleine Heyder[2], Karen Forberich[1,2], Larry Lüer[1,2], Christoph J. Brabec[1,2], H.-J. Egelhaaf[1,2]

[1] Forschungszentrum Jülich GmbH, Helmholtz-Institute Erlangen-Nürnberg for Renewable Energy (HI ERN), Dept. of High Throughput Methods in Photovoltaics, Erlangen, Germany

[2] Friedrich-Alexander-Universität Erlangen-Nürnberg, Materials for Electronics and Energy Technology (i-MEET), Erlangen, Germany

**1. Spectral Data Processing and Analysis:**

Due to inhomogeneities in the active layer (slightly fluctuating thickness) and the movement of the web (slight variance of web position and tilt with respect to the spectrometer beam), the spectral raw data has a noisy appearance. However, the majority of this "noise" is systematic and identical for all wavelengths, as can be seen in Figure S1 (grey curves). Consequently, the temporal evolution of the absorption at the isosbestic point (~600 nm), i.e., the wavelength at which the absorbance should – in theory – stay constant over time, can be used to correct/smoothen the signals at all other wavelengths using Equation S1.

$$A_{corrected}(\lambda, t_x) = \frac{A_{raw}(\lambda, t_x)}{A_{isosbestic}(t_x)} * \bar{A}_{isosbestic} \qquad \text{Eq. S1}$$

Here, $A_{raw}(\lambda, t_x)$ is the absorbance at a certain wavelength and a certain point in time, $A_{isosbestic}(t_x)$ the respective absorbance at the isosbestic point, and $\bar{A}_{isosbestic}$ the average absorbance over all times at the isosbestic wavelength. The resulting processed data is also plotted in Figure S1 (colored curves).

The respective mass fractions of the donor and acceptor are calculated from the spectral data according to Equation S2 and Equation S3.

$$w_D(t_x) = \frac{A_{corrected}(517\,nm, t_x) - A_{corrected}(517\,nm, t_0)}{A_{corrected}(517\,nm, t_{end}) - A_{corrected}(517\,nm, t_0)} \qquad \text{Eq. S2}$$

$$w_A(t_x) = 1 - w_D(t_x) \qquad \text{Eq. S3}$$

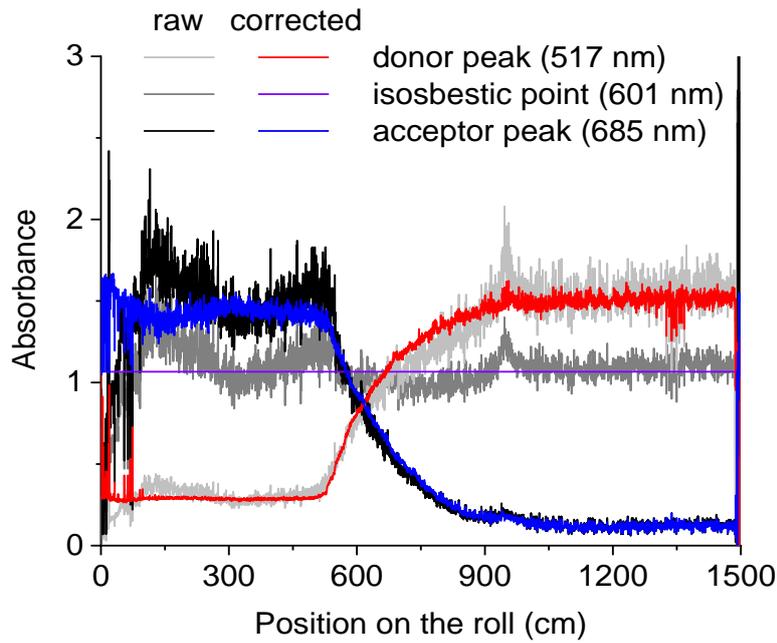

Figure S1: Temporal evolution of the spectral raw data (grey/black) and the data processed according to Eq. S1 (colored) for three different wavelengths.

## 2. Determination of dry film thickness:

The dry film thickness of the $SnO_2$ variation is calculated with the following formula, where we assumed that the space filling of $SnO_2$ nanoparticles is 0.74. The bulk densities of $SnO_2$ and Butanol are 6.95 g/cm³ and 0.81 g/cm³, respectively. This calculation gives a rough estimation of the final dry film.

$$Dry\ film\ thickness = \frac{Flowrate}{Web\ speed \times Coating\ width} \times \frac{Conc(w/w) \times \rho\ BuOH}{\rho\ SnO_2 \times packing\ density} \quad \text{Eq. S4}$$

## 3. Experimental Design:

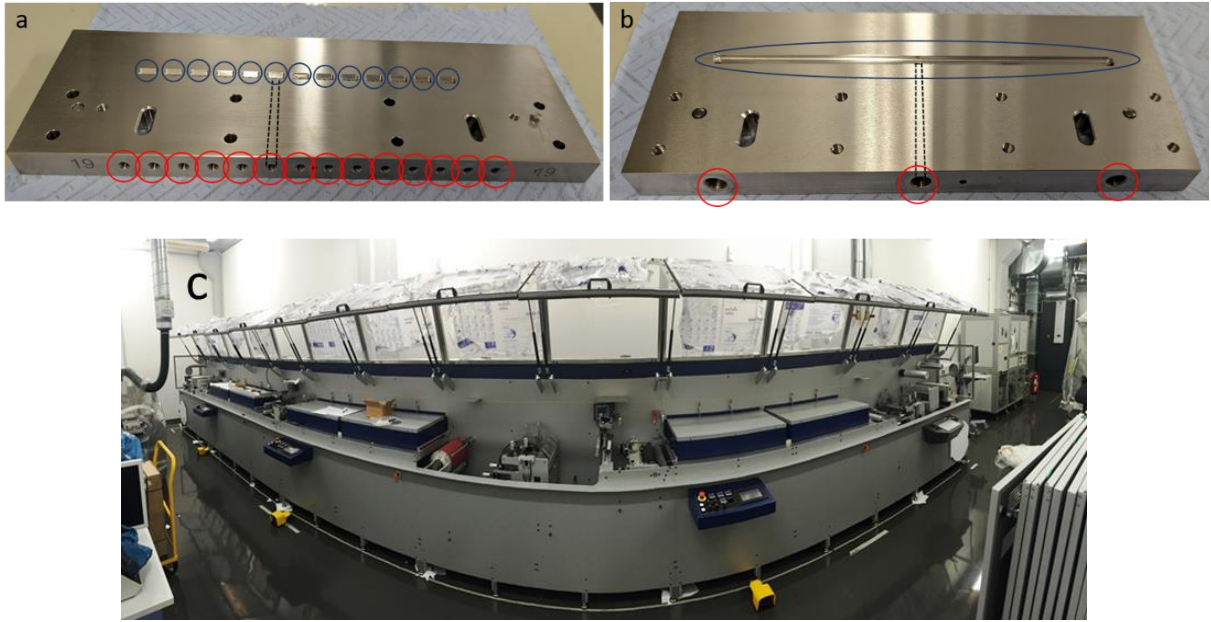

Figure S2: (a) Multi Channel Slot Die with 13 individual inlets (red circles) and reservoirs (blue circles) which is used for SnO$_2$ coating. (b) Slot Die with one reservoir (blue oval) and three inlets/outlets (red circles), which is used for active layer, HTL and Ag NW coating. On each photograph, one of the channels which connect inlets to reservoirs is marked exemplarily with black dashed lines. (c) full view of the roll-to-roll coating machine.

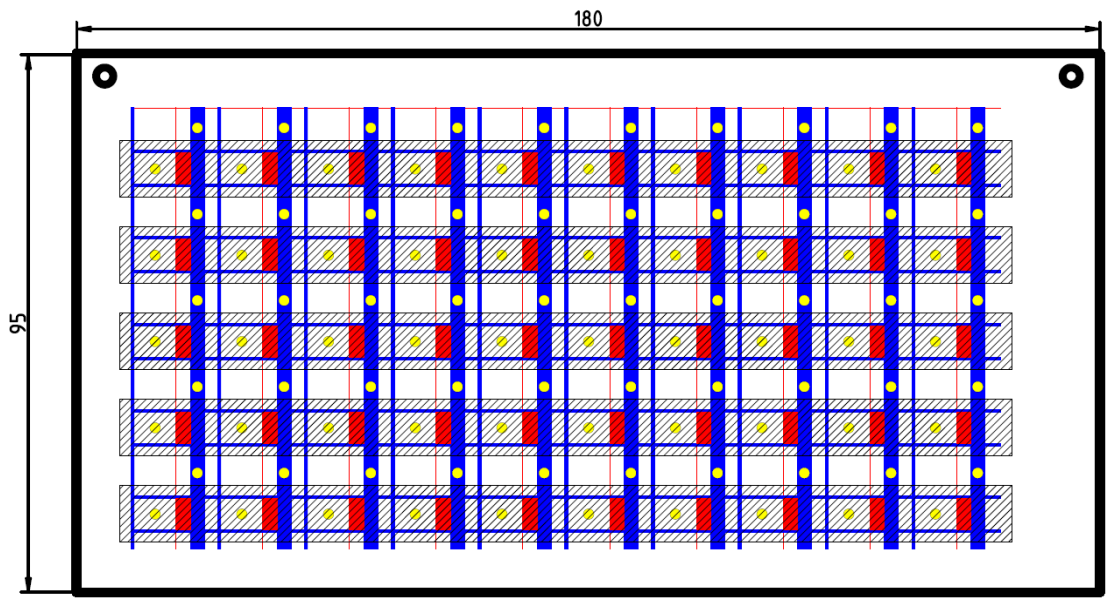

Figure S3: Schematic of the measured sheets, where the red lines mark the P1 lines (removal of bottom electrode), the blue lines the removal of the top electrode, the red boxes the active area of the solar cells, the dashed lines the five different $SnO_2$ stripes, the yellow circles the contact spots of the top and bottom electrode and the black circles the alignment holes.

## 4. GPR Analysis

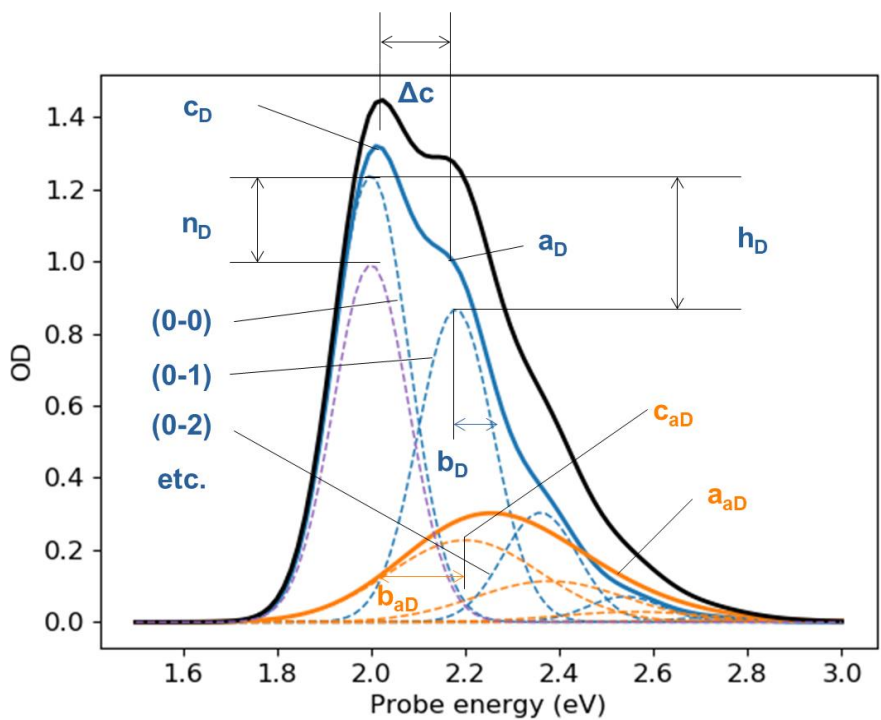

Figure S4. Spectral decomposition of the lowest energetic optical absorption of the donor polymer (suffix "D") in the solid state into contributions from the ordered (blue) and amorphous phase (orange, suffix "a"). The decomposed spectra are further decomposed into vibronic contributions, according to Spano's model [Refs. XXIII, XXIVa (main manuscript)], assuming a single essential vibronic progression (dashed lines). For a donor-acceptor blend, the same analysis is done for the lowest energetic acceptor absorption. As the relative height of the dashed lines and their energetic spacing are both given by a single parameter ($h_D$ and $\Delta c$, respectively), the model accommodates significant spectral overlap between donor and acceptor absorption while still yielding reasonable uncertainties for the phase specific morphological parameters

**List of spectral features and their relation to morphological parameters**

| Name [unit] | Provenience | Meaning | Morphology relation |
|---|---|---|---|
| $a_{D/A}$ [eV] | Fit parameter | Total area under blue curve | Persistence length, anisotropy, film thickness |
| $b_{D/A}$ [eV] | Fit parameter | Gaussian bandwidth of each single blue dashed curve | Energetic disorder |
| $\underline{c_{D/A}}$ [eV] | Fit parameter | Center energy of dashed curve for (0-0) vibronic transition | Domain size / dielectric coupling |
| $h_{D/A}$ [] | Fit parameter | Huang-Rhys factor for single effective vibronic progression | Wavefunction delocalization (on-chain ordering) |
| $n_{D/A}$ [] | Fit parameter | Relative suppression of (0-0) vibronic due to weak H aggregation according to Spano's model (fixed to 0.5) | Weak H aggregates (on-chain ordering) |
| $\Delta c$ [eV] | Fit parameter | Effective single vibronic progression (fixed to 0.185 eV) | Vibronic coupling |
| $a_{tot}$ [eV] | $a_D + a_{aD} + a_A + a_{1A}$ | Total area under black curve (donor + acceptor) | Film thickness, anisotropy |
| $A_{tot}$ [eV] | $a_A + a_{aA}$ | Total area under black curve (only acceptor) | Total amount of acceptor |
| $D_{tot}$ [eV] | $a_D + a_{aD}$ | Total area under black curve (only acceptor) | Total amount of donor |
| $X_A$ [] | $A_{tot}/a_{tot}$ | Relative spectral weight of acceptor in lowest energetic optical transition | D:A ratio |
| $X_{amD}$ [] | $a_{aD}/D_{tot}$ | Relative spectral weight of amorphous phase in donor absorption | Order, domain size |
| $X_{amA}$ [] | $a_{aA}/A_{tot}$ | Relative spectral weight of amorphous phase in acceptor absorption | Order, domain size |

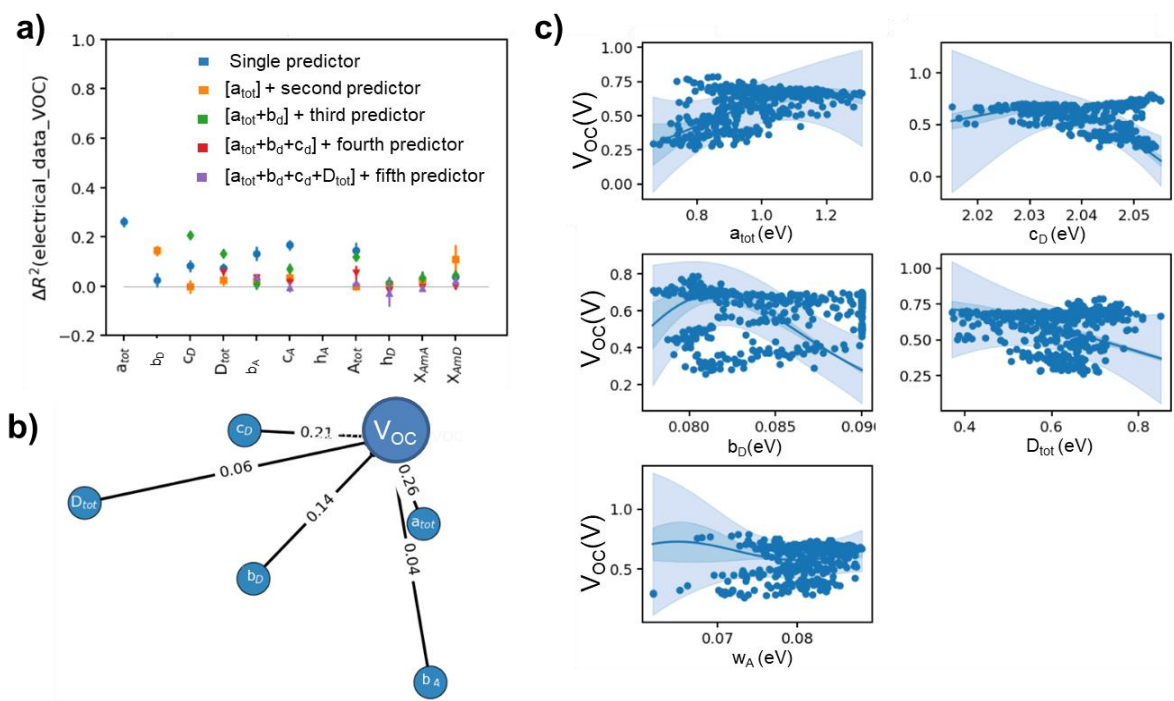

Figure S5: Prediction of $V_{OC}$ from active layer morphology by minimum Redundancy Maximum Relevance embedded Gaussian Process Regression (mRMR-GPR). a) results mRMR-GPR runs: single predictor (blue symbols); two predictors including best blue predictor (orange); three predictors

including best blue and best orange predictor (green), four predictors (red), five predictors(purple). b) differential explanation of variance by each selected predictor. Explanation values below 10% are considered irrelevant. The predictors are arranged in a spring model according to their explanation of variance c) one-dimensional intersections through the approximate objective function for $V_{OC}$, as obtained by mRMR-GPR (blue solid line), light blue and dark blue areas: 95% confidence intervals for the uncertainty of a single prediction and uncertainty of the mean, respectively.

Figures S5a shows the working principle of the minimum Redundancy Maximum Relevance (mRMR) embedded feature extraction using GPR [Ref. XXII (main manuscript)]. First, $V_{OC}$ is predicted using only single predictors one by one (blue symbols). Several features are able to explain around 20 % of the variance in the measured $V_{OC}$ values. Next, the strongest single predictor ($a_{tot}$) is retained and a second predictor is included from the remaining predictor list (orange symbols). Comparing the orange symbols with the blue ones, we find for most of the predictors (see e.g. $b_A$, $c_A$) that they do not provide additional explanation of variance once $a_{tot}$ is considered. However, there are two examples ($b_D$ and $X_{amD}$) where the explanation of variance increases under the presence of $a_{tot}$ (orange symbol has higher value than blue symbol). This points to a non-linear correlation between $b_D$ and $V_{OC}$ which was only found once the main correlation of $a_{tot}$ was included in the feature list. Again, the stronger of these two predictors($b_D$) is included into the feature list, and the procedure is repeated (green symbols) until a maximum number of allowed features is reached (5 in this work).

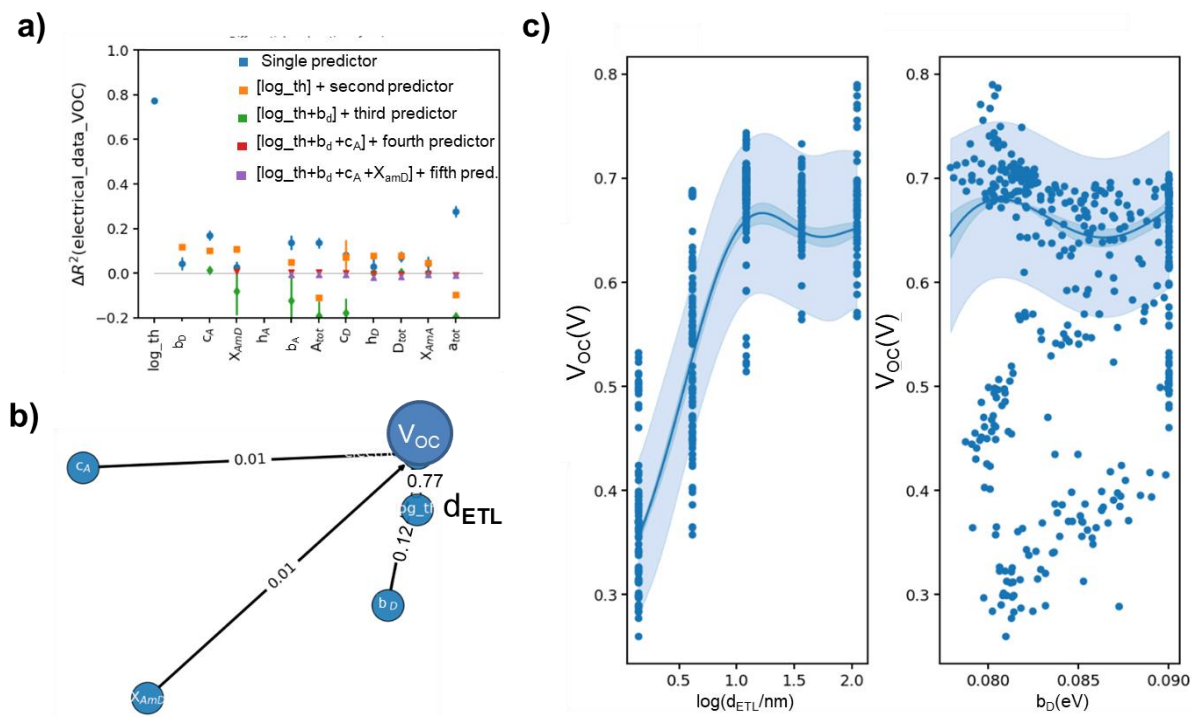

Figure S6: Prediction of $V_{OC}$ from active layer morphology and $d_{ETL}$ by mRMR-GPR. a) results of mRMR-GPR runs clearly showing that $d_{ETL}$ is more important for $V_{OC}$ than AL morphology: as a single predictor

(blue), $a_{tot}$ explains 25% of the variance, as shown in Figure S2a. But as soon as $d_{ETL}$ (given as logarithmic value named **log_th**) is included into GPR (orange), $a_{tot}$ provides no additional explanation of variance, compare blue symbol with symbols of other colors at position "$a_{tot}$", b) explanation of variance by selected predictors, c) one-dimensional intersections through the approximate objective function for $V_{OC}$, as obtained by mRMR-GPR (blue solid line), light blue and dark blue areas: 95% confidence intervals for the uncertainty of a single prediction and uncertainty of the mean, respectively.

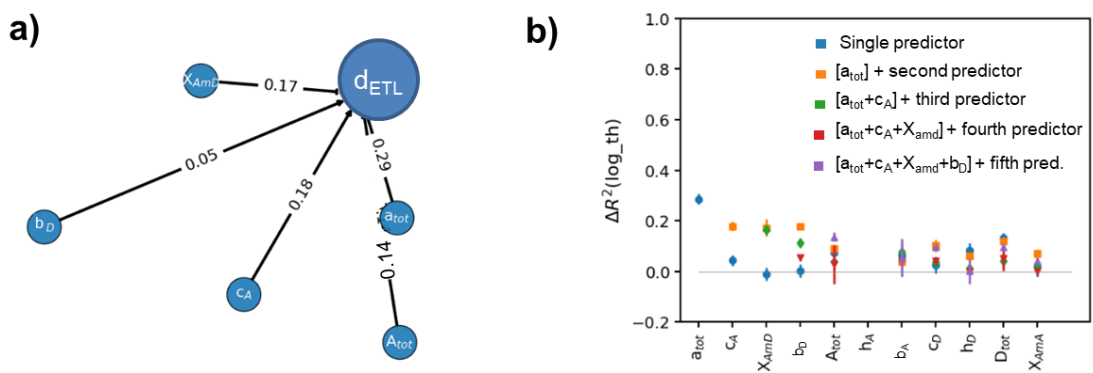

Figure S7: Prediction of $d_{ETL}$ from active layer morphology by mRMR-GPR. a) Differential explanation of variance by selected predictors, b) mRMR – GPR procedure.

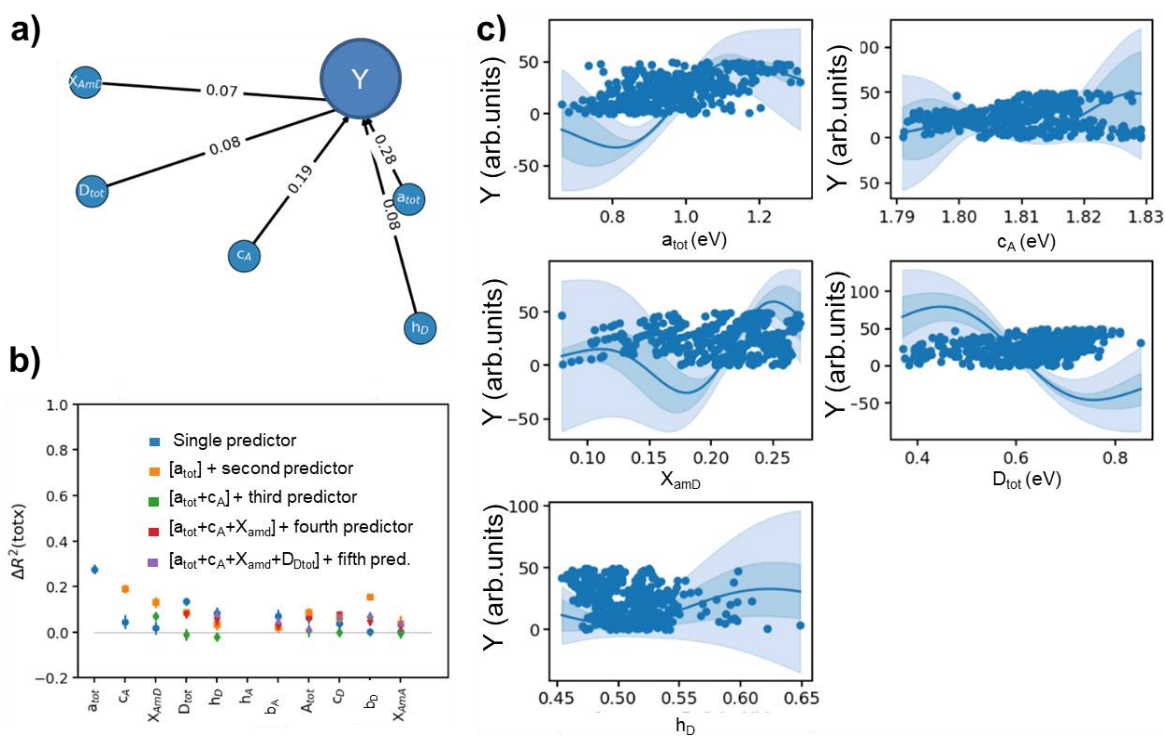

Figure S8: Prediction of the cross-web position Y from active layer morphology by mRMR-GPR. a) Differential explanation of variance by selected predictors, b) results of mRMR – GPR procedure. c) One-dimensions intersections through the approximate objective function for $V_{OC}$, as found by GPR

(blue solid line), light blue and dark blue areas: 95% confidence intervals for the uncertainty of a single prediction and uncertainty of the mean, respectively.

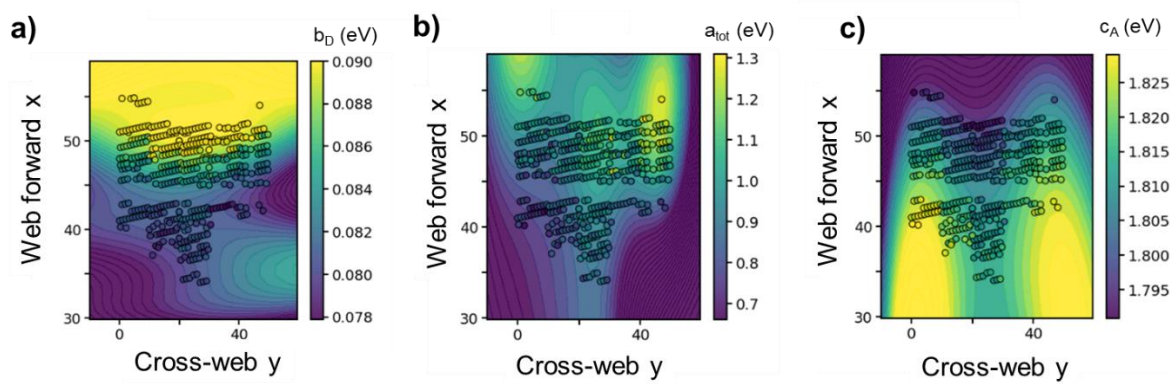

Figure S9. Evolution of active layer morphology across (y) and along (x) the web direction. a) Donor exciton bandwidth, b) total exciton absorption (donor + acceptor), c) acceptor exciton energy.

Figure S9 shows the evolution of active layer morphology across (y) and along (x) the web direction. We find that the donor bandwidth (panel a) shows a variation nearly exclusively along x, where the D:A ratio is varied. Hence, in acceptor-rich blends (x>50, corresponding to $w_A \cong 0.7 - 0.75$), we find that donor bandwidth is high, which may point to an increase of disorder in the donor phase. From this observation, in Fig. 5d in the main text we draw causal links from x to $w_A$, and from $w_A$ to $b_D$, but we draw no arrow from y to $b_D$.

In contrast, the total exciton absorption $a_{tot}$ is mainly influenced along the cross-web direction y, see Fig. S8b. The total absorption, and hence the film thickness, is especially high for y>40, which is the region of the thinnest ETL stripe. This explains the high correlation between ETL thickness and $a_{tot}$, found in Fig. S4a; therefore, in the knowledge graph in Fig. 5d, we can draw a causal connection from y to an unknown processing condition $P_{AL}$ acting on AL, and from $P_{AL}$ to $a_{tot}$. We can speculate that $P_{AL}$ is given by an unequal distribution of nozzle flow rates.

Finally, the acceptor exciton energy $c_A$ is influenced by both x and y, see Fig. S7c. A lower exciton energy means extended J aggregates. Along the web forward direction, we find that the exciton energy decreases for acceptor-rich blends, which makes sense because domains will be the larger the less disturbed by the polymer donor. However, we also see a symmetric evolution of $c_A$ along the cross-web direction y, meaning that the exciton energy is higher at the edges than in the center. This means that at constant $w_A$, larger acceptor domains are formed in the center than in the edges, which may come from a small temperature gradient in the annealing oven, being slightly colder in the edges.

## 5 Optimal processing parameters

In order to find the uncertainty of the optimal processing parameters, we perform a brute force sampling of the two-dimensional objective function displayed in Fig. 4 b/c to encounter the data range where $u_{cb}$ >= max(PCE), where $u_{cb}$ is the upper confidence bound for PCE with respect to the uncertainty of the mean. We find $w_{A,i}$=0.59+-0.04; and **$d_{ETL}$** = 17+-4 nm.

## 6. Drift-Diffusion Simulations

The drift-diffusion simulations were performed using the open-source program SIMsalabim version 4.45.[1, Ref. XXV in main manuscript]

SIMsalabim solves the 1D drift-diffusion equations which consist of a set of three main equations, the Poisson, continuity and drift-diffusion equations.

The Poisson equation:

$$\frac{\partial}{\partial x}\left(\varepsilon(x)\frac{\partial V}{\partial x}\right) = -q\,(p(x) - n(x) + C_i(x))$$

where *x* is the position in the device,[xxv] *q* the electric charge, *V* the electrostatic potential, *n* and *p* the electron and hole concentrations, and *ε* the permittivity. $C_i$ can represent any other type of charges in the systems such as: (i) doping with $N^-_A$ and $N^+_D$ being the ionized p-type and n-type doping respectively, or (ii) the charged traps $\Sigma^+_T$ and $\Sigma^-_T$ for hole and electron traps. Such as the Poisson equation may be written as:

$$\frac{\partial}{\partial x}\left(\varepsilon\frac{\partial V}{\partial x}\right) = -q\,(p - n + N^-_D - N^+_A + \Sigma^+_T - \Sigma^-_T))$$

The current continuity equations:

$$\frac{\partial J_n}{\partial x} = -q\,(G - R)$$

$$\frac{\partial J_p}{\partial x} = q\,(G - R)$$

with $J_{n,p}$ the electron and hole currents, *G* and *R* the generation and recombination rate respectively.

The movement of these free charges is governed either by diffusion due to a gradient in carrier density or by drift following the electric field such as the electron and hole currents can be written as:

$$J_n = -q\,n\,\mu_n\,\frac{\partial V}{\partial x} + q\,D_n\,\frac{\partial n}{\partial x}$$

$$J_p = -q\,p\,\mu_p\,\frac{\partial V}{\partial x} - q\,D_p\,\frac{\partial p}{\partial x}$$

with $\mu_{n,p}$ the charge carrier mobilities and $D_{n,p}$ carrier diffusion coefficients. The carrier diffusion coefficients can be written following Einstein's equation such as:

$$D_i = \frac{k_B T}{q} \mu_i$$

with $k_B$ the Boltzmann's constant, $T$ the absolute temperature.

For the simulation, we chose to place the cathode at $x = 0$ and the anode at $x = L$ as a convention, $L$ being the total thickness of the device.

In order to numerically solve the system of equations presented above we need to specify the boundary conditions for the carrier densities:

$$n(0) = N_c \, exp\left(-q\frac{\varphi_n}{k_B T}\right) \quad \cdots \quad n(L) = N_c \, exp\left(-q\frac{E_g - \varphi_n}{k_B T}\right)$$

$$p(0) = N_v \, exp\left(-q\frac{E_g - \varphi_p}{k_B T}\right) \quad \cdots \quad p(L) = N_v \, exp\left(-q\frac{\varphi_n}{k_B T}\right)$$

and the potential at the contacts:

$$q\left(V(L) - V(0) + V_{app}\right) = W_c - W_a$$

with $N_c$ and $N_v$ the effective density of states for the conduction and valence band respectively, here we chose $N_c$ and $N_v$ to be equal, $\varphi_n$ and $\varphi_p$ the electron and hole injection barrier at the cathode and anode, $V_{app}$ being the externally applied voltage and $W_a$ and $W_c$ the anode and cathode work functions respectively.

The recombination rate $R$ is typically expressed by adding the contribution from the band-to-band/bimolecular recombination and Shockley-Read-Hall (SRH) recombination from equations:

$$R_B = \gamma(np - n_i^2)$$

$$R_{SRH} = \frac{C_n C_p \Sigma_T}{C_n(n + n_1) + C_p(p + p_1)} (np - n_i^2)$$

$\gamma$ is the bimolecular recombination rate constant, $n_i$ is the intrinsic carrier concentration, $n_1$ and $p_1$ are constants that depend on the trap energy level ($E_{trap}$), and $C_n$ and $C_p$ are the capture coefficients for electrons and holes respectively. $n_1$ and $p_1$ are defined as followed:

$$n_1 = N_c \exp\left(-q\frac{E_C - E_{trap}}{k_B T}\right)$$

$$p_1 = N_v \exp\left(-q\frac{E_{trap} - E_v}{k_B T}\right)$$

For more information about how this system of equations is solved we encourage the readers to read references [2,3,4].

Table S1: Device parameters used to simulate the devices. The varied parameters values for each donor:acceptor combination and the different $SnO_2$ thicknesses can be found in Figures S10a-f. Note that for simplicity and to remove some parameters to optimize for the fits the PEDOT:PSS/AgNW is set as an effective electrode forming an ohmic contact with the active layer and with a work function $W_a$.

| Parameter | Unit | Value |
|---|---|---|
| P3HT:o-IDTBR | | |
| $E_c$ / $E_v$ | eV | 3.6 / 4.8 |
| $N_c$ | $m^{-3}$ | $2 \times 10^{27}$ |
| L | nm | 100 |
| $\varepsilon_r$ | - | 3.5 |
| $\mu_p$ | $m^2\,V^{-1}\,s^{-1}$ | $2.52 \times 10^{-8}$ |
| $\mu_n$ | $m^2\,V^{-1}\,s^{-1}$ | Varied |
| $G_{ehp}$ | $m^{-3}\,s^{-1}$ | Varied |
| $\gamma$ | $m^3\,s^{-1}$ | Varied |
| $\Sigma_T$ | $m^{-3}$ | Varied |
| $E_{trap}$ | eV | 4.21 |
| $C_n$ / $C_p$ | $m^3\,s^{-1}$ | $10^{-13}$ |
| $SnO_2$ | | |
| $E_c$ / $E_v$ | eV | Varied / 8.18 |
| $N_c$ | $m^{-3}$ | $3.63 \times 10^{24}$ |
| L | nm | 4.1 – 12.2 – 36.6 |
| $\varepsilon_r$ | - | 10 |
| $S_T$ ($SnO_2$ Interface) | $m^{-2}$ | Varied |
| Electrode Work functions | | |
| $W_a$ (PEDOT:PSS/AgNW) | eV | 4.8 |
| $W_c$ (IMI) | eV | Aligned to the $SnO_2$ $E_c$ |
| External Parameters | | |
| $R_s$ | $\Omega\,m^2$ | Varied |
| $R_{sh}$ | $\Omega\,m^2$ | Varied |
| T | K | 295 |

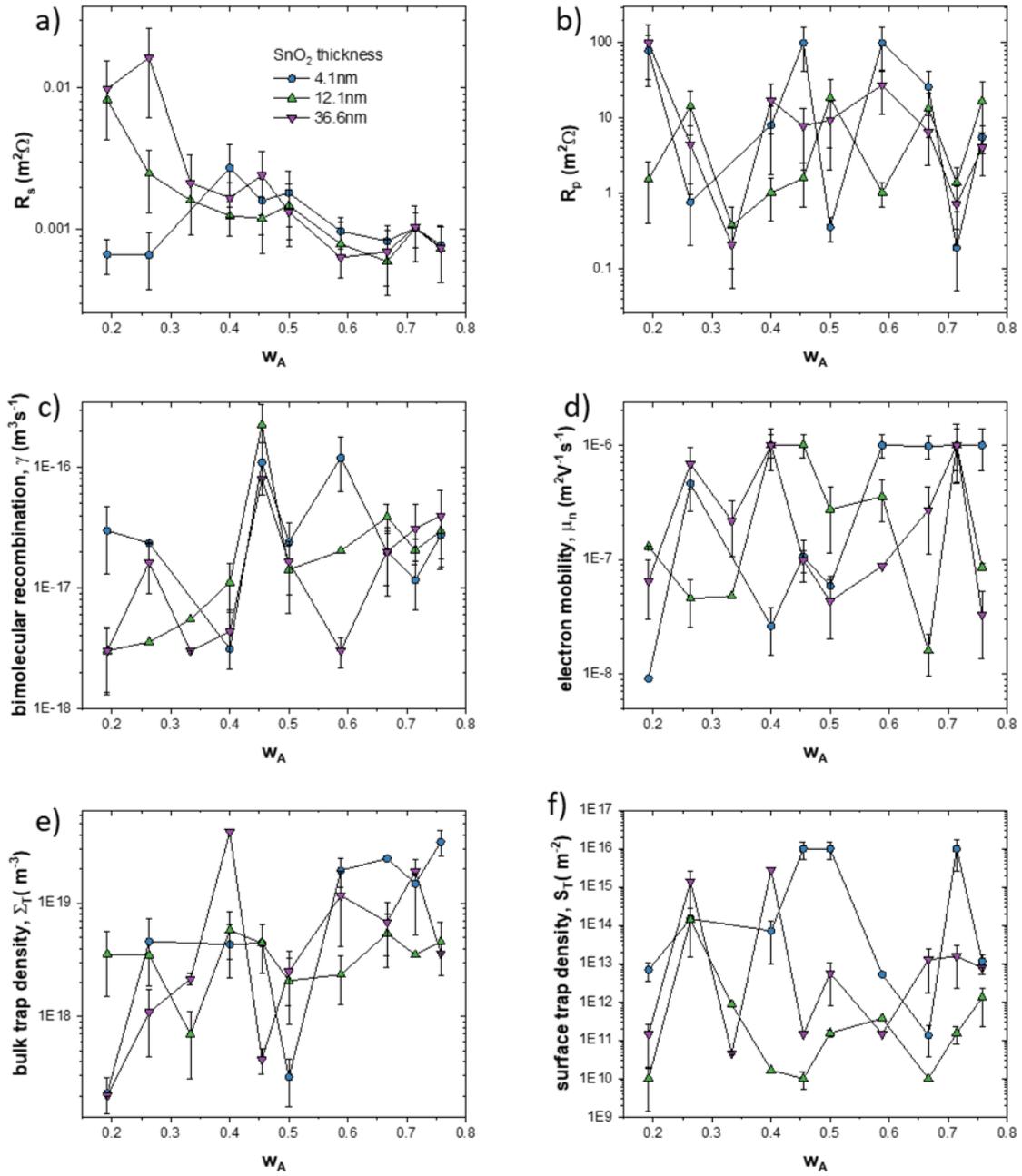

Figure S10: Additional parameters obtained from drift-diffusion fitting, plotted against the acceptor mass fraction $w_A$ and for different ETL thicknesses (blue: 4.1nm, green: 12.1nm, violet: 36.6 nm, as given in panel a). a) series resistance $R_s$, b) parallel resistance $R_p$, c) bimolecular recombination coefficient $\gamma$, d) electron mobility $\mu_n$, e) bulk trap density $\Sigma_T$, f) surface trap density $S_T$. None of these parameters show a clear dependence on the acceptor fraction or the ETL thickness.

## 7. Fitting procedure with Bayesian optimization

Our fitting procedure is based on the bayesian optimization package from scikit-optimize[https://scikit-optimize.github.io/stable/] using the skopt.Optimizer framework.

The skopt.Optimizer is used to minimize the mean-square error (MSE) between the experimental data and the simulated data with SIMsalabim by optimizing the value of the different material parameters described in table S1. Figure S11 describes the logic behind the optimization procedure. In our case, the experimental data is the 1 sun JV curve for each device and the physical model is the drift-diffusion model described in the previous section. The cost function to minimize is the MSE and we used a gaussian process regressor (GPR) as a surrogate model (we also tested other surrogates but the GPR performed the best). To ensure a good balance between exploration and exploitation of the entire parameter space we used the 'gp_hedge' option for the acquisition function. We performed a random initial sampling (80 points) using the Latin hypercube algorithm followed by 200 points of bayesian optimization distributed over 4 cores.

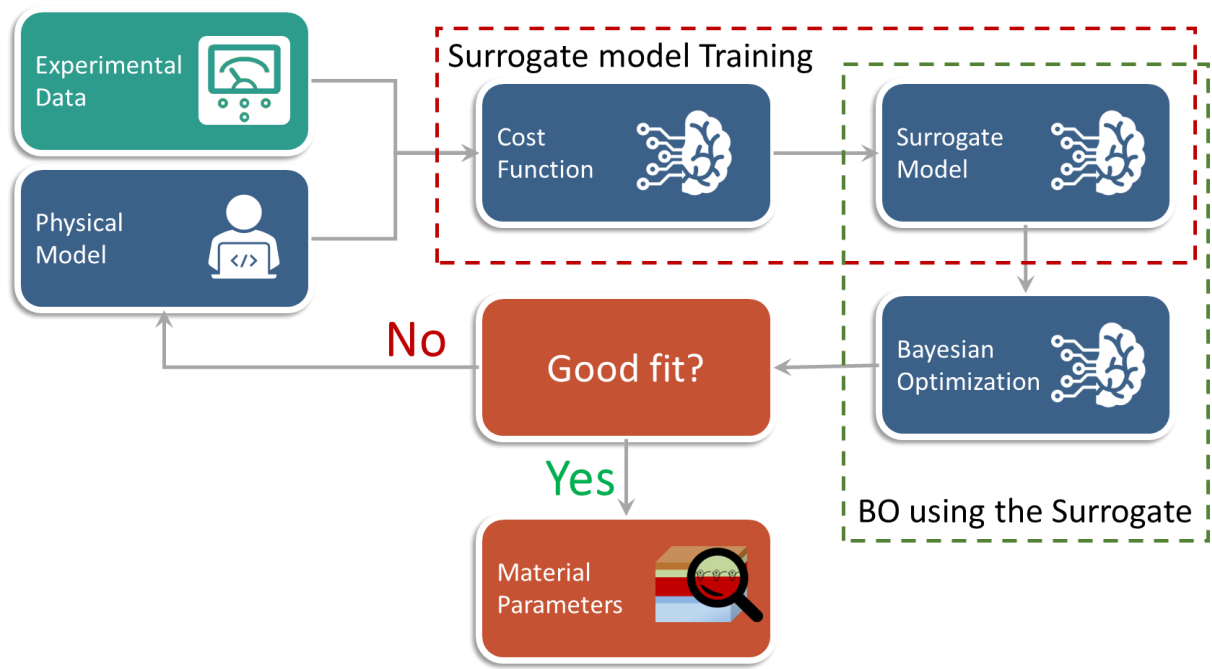

Figure S11: Flowchart describing the optimization procedure to perform the JV-curve fitting procedure using Bayesian optimization.

References

[1] Koopmans et al., (2022). SIMsalabim: An open-source drift-diffusion simulator for semiconductor devices. Journal of Open Source Software, 7(70), 3727, https://doi.org/10.21105/joss.03727

[2] Selberherr, S., Analysis and Simulation of Semiconductor Devices, Springer (1984)

[3] Koster, L.J.A., Smits, E.C.P., Mihailetchi, V.D., Blom, P.W.M., Device model for the operation of polymer/fullerene bulk heterojunction solar cells, *Phys. Rev. B* **72**, 085205 (2005)

[4] Koopmans, M., Koster, L.J.A., Voltage Deficit in Wide Bandgap Perovskite Solar Cells: The Role of Traps, Band Energies, and Effective Density of States, Solar RRL 6, 2200560 (2022)